# Point contact Andreev reflection spectroscopy of NdFeAsO$_{0.85}$

K A Yates[1,3], L F Cohen[1], Zhi-An Ren[2], Jie Yang[2], Wei Lu[2], Xiao-Li Dong[2] and Zhong-Xian Zhao[2]

[1] The Blackett Laboratory, Physics Department, Imperial College London, SW7 2AZ, UK

[2] National Laboratory for Superconductivity, Institute of Physics and Beijing National Laboratory for Condensed Matter Physics, Chinese Academy of Sciences, PO Box 603, Beijing 100190, PR China

The newly discovered oxypnictide family of superconductors show very high critical temperatures of up to 55K. Whilst there is growing evidence that suggests a nodal order parameter, point contact Andreev reflection spectroscopy can provide crucial information such as the gap value and possibly the number of energy gaps involved. For the oxygen deficient NdFeAsO$_{0.85}$ with a T$_c$ of 45.5K, we show that there is clearly a gap value at 4.2K that is of the order of 7meV, consistent with previous studies on oxypnictides with lower T$_c$. Additionally, taking the spectra as a function of gold tip contact pressure reveals important changes in the spectra which may be indicative of more complex physics underlying this structure.

[3] Corresponding author, Karen A Yates, e-mail: k.yates@imperial.ac.uk

Earlier this year, the oxypnictide family of superconductors were found to have transition temperatures of 26K[1]. It was quickly established that physical pressure[2] or chemical doping[3,4] could enhance the critical temperature to above 40K. Various theories have been proposed to explain the high transition temperatures observed[5,6,7,8,9] and as raised in those papers several key questions need to be asked. Establishing the symmetry of the order parameter, whether multiple superconducting gaps play a role and whether the magnitude of the gap value(s) scale with the $T_c$ of the material. Regarding the first of these questions, it is rapidly becoming apparent that a nodal order parameter may be needed to explain the experimental data[10,11,12,13,14]. For the other questions, a technique that probes the gap structure such as scanning tunnelling microscopy (STM) or point contact Andreev reflection (PCAR) spectroscopy can prove very insightful. A previous PCAR study on $LaO_{0.9}F_{0.1-\delta}FeAs$, ($T_c \approx 26K$) indicated a maximum value for the superconducting energy gap of $\Delta_0 \approx 3.9 \pm 0.7 meV$[10]. Furthermore, the spectra in that study could only apparently be fitted assuming theoretical models incorporating a p-wave or d-wave symmetry. In contrast $NdFeAsO_{0.85}$ has a much higher $T_c$[3] and, unlike the early oxypnictides carriers are not created by fluorine doping but by oxygen deficiency. Density functional theory calculations have indicated that both oxygen vacancies and F doping act in a similar way to introduce carriers into the FeAs layers[15] although Mössbauer studies have indicated that there are differences in the magnetic environment experienced by the Fe ions for oxygen deficient samples compared with F doping[16,17]. So many of these questions remain open. Experimentally, the influence of the oxygen stoichiometry on the superconducting properties of the material has been studied by Ren et al.,[3] where they showed a dome shaped phase diagram of $T_c$ vs oxygen stoichiometry. In this sense, the oxypnictide family appears highly reminiscent of the cuprate based high temperature superconductors (HTS).

The symmetry of the order parameter has a profound effect on the conductance spectra measured by PCAR. For a simple s-wave order parameter, the conductance below the superconducting gap voltage ($\Delta$) is enhanced by Andreev reflection and the conductance peaks at $\pm\Delta$ leading to a minimum at zero bias. Such spectra can be fitted by the Blonder Tinkham Klapwijk (BTK) model in order to determine the value of $\Delta$[18]. For a d-wave order parameter, the presence of nodes in the order parameter over the Fermi surface leads to Andreev bound states (ABS) at the surface of the material which manifest as peaks at zero bias in the PCAR conductance spectra, for a review see [19]. Although the presence of a zero bias conductance peak (zbcp) is in many ways the signature of a nodal order parameter, other factors such as Josephson effects[10,23] must be ruled out as potential causes of the zbcp.

The $NdFeAsO_{0.85}$ sample studied here was prepared by high pressure synthesis as described elsewhere[3,4]. The resistivity was measured by a four point technique and by defining $T_c^{on}$ as 90% $R_n$, a $T_c^{on}$ of 45.5K was determined for this sample. The width of the transition was $\Delta T_c$ = 6.8K (90%-10%). Point contact Andreev reflection spectra were obtained using a mechanically sharpened Au tip as previously reported[20]. A differential screw mechanism was used to establish a point contact that could be measured as a function of tip pressure and temperature. In order to measure the four terminal differential conductance, a DC bias was applied across the tip-sample contact and an ac ripple applied.

Figure 1 shows the PCAR spectra at various temperatures. At 5K and 10K temperatures, there is a pronounced zbcp with well defined shoulders on the spectra between 5-7meV, as well as multiple weaker peak features at higher bias. As the temperature is increased, these features move to lower bias and become increasingly smeared.

Close to $T_c$, the spectra shown in figure 1 can be normalised to the conductance spectrum just above $T_c$. In order to track the behaviour of the zbcp as a function of temperature, the magnitude of the zbcp normalised in this way is shown in figure 1b as a function of temperature for two contacts made at two different locations on the sample. When normalising over a wider temperature range, caution must be exercised due to the possible temperature dependence of the normal state[21]. With this in mind, only data taken within 15K of $T_c$ are shown in figure 1b. For both contacts there is no indication of superconductivity in the PCAR spectra above 46K. However, as the temperature is reduced below 42K (contact 1) and 45K (contact 2) a zbcp appears which increases in magnitude with decreasing temperature. The difference between the temperature of onset of the two data sets is consistent with the samples having a broad superconducting transition i.e sample inhomogeneity and the fact that the macroscopic tip dimension was of the order of 20-50μm resulting in different contacts having slightly different $T_c$ values. Consistent with the previous study[10], the magnitude of this peak increases sharply with decreasing temperature below $T_c$ indicating that the peak is associated with the material once it is superconducting although it is not absolutely clear that its existence is due to the presence of a nodal order parameter as in [10] or whether it is related to Josephson junctions, such as those that may exist in series with the contact due to the polycrystalline nature of the sample[23].

In the HTS materials the magnitude of the zbcp is known to vary both with crystalline orientation and effective barrier potential, $Z$[19]. For most directions a zbcp is observed that increases in magnitude with increasing $Z$[22]. The size of the zbcp is determined by the angle, α, which the tip makes relative to the order parameter in the superconductor. So it is plausible that at certain angles the peak may not be observed. As the sample here is polycrystalline we can expect to average over several grains with random orientation, therefore we would expect to see a zbcp in most spectra, though in some it may be quite broad[23]. We do in fact see many spectra where the peak is absent. Note that the original BTK model implicitly assumes a nodeless superconducting gap and so for all crystal orientations in a superconductor having a nodal order parameter, the PCAR spectra would need to be fitted by the Tanaka and Kashiwaya[22] adaptation of the BTK model[18].

A series of PCAR spectra was collected as a function of tip pressure and the results are shown in figure 2. As the contact is pushed into the sample, it is likely that it punctures through oxide layers that exist on the surface[24]. This has the effect of decreasing the effective barrier height, $Z$[20]. Measuring the conductance as a function of tip pressure over several different contacts over the surface of the bulk therefore provides information both of the evolution of the zbcp with Z as well as on sample homogeneity. Two types of contact were observed as illustrated by figures 2(a) and (b). Spectra such as those in figure 2(a) had a zbcp that was visible at low contact pressure (high resistance), with shoulder features around ±7meV, that became a zero bias conductance minimum at higher tip pressures (low contact resistance). As has been pointed out by Biswas et al.,[25] in the case of $Pr_{2-x}Ce_xCuO_4$ this high tip pressure spectrum is indicative of a nodeless order parameter. Following the suggestion of Biswas et al., we also estimate a gap from the second spectrum in figure 2(a) of $\Delta_0 \sim 7$meV. This gives a value of $2\Delta_0/k_BT_c = 3.57$ compared with $2\Delta_0/k_BT_c \sim 3.5$ from the results of Shan et al[10]. Although this result is tentative given that full spectral fitting using the Tanaka and Kashiwaya model would be needed to check the precise $\Delta$ values, the similarity between the two point contact results for materials with different stoichiometry shows that similar pairing mechanisms are likely to be involved in these samples. The correspondence with the weak electron - phonon coupling BCS theory expression is also very interesting. Our results are also in accordance with Sato et al.,[26] who have found $2\Delta_0/k_BT_c$ of ~4 for the oxpnictides, although they contrast with the results of Dubroka et al[27] who obtain a value of $2\Delta_0/k_BT_c = 8$. The discussion of whether these relationships indicate that a symmetric BCS weak coupling limit is an appropriate description would be premature given the nature of the samples as well as the apparent conflict with the observation of the zero bias conductance peak.

Spectra such as those in figure 2(b) show that at low tip pressure there is a zero bias conductance dip and as the tip pressure is increased (contact resistance decreases), a small zero bias conductance peak develops. It is possible that the higher contact resistance contacts have a greater amount of non-thermal smearing[20], and the zbcp may be obscured under these circumstances or it may be that the oxide layers at the surface are magnetic and some processes are suppressed until the oxide is punctured. The low tip pressure results may result from a proximity effect between the superconductor and this surface layer[28]. An example of a clear low tip pressure contact similar to that shown in figure 2(b) is shown in figure 3. This spectrum can be fit very well with the original BTK model (assuming a nodeless order parameter). The BTK fit gives an extracted gap value of 5.94meV, slightly lower than the ~7meV indicated by the spectra in figure 2(a). A lower gap value supports the possibility of proximity effects at the surface of these samples. The other feature of note in figure 2(b) is that these spectra show a pronounced asymmetry, which is consistent with both the PCAR spectra for LaFeAsOF and theoretical predictions[7].

The fact that such a good fit is obtained from some of the spectra with a model developed for simple s-wave symmetry is indicative of the fact that, as with the electron doped cuprates the understanding of the order parameter in the oxypnictides may be far from simple. Indeed the existence of an order parameter that can be fitted by the BTK model does not necessarily imply that this material is a conventional s-wave superconductor. Unconventional s-wave superconductivity of the form proposed by Mazin et al., would also lead to such a result[5]. Furthermore, results such as these have also been observed in the electron doped cuprates and can result from proximity effects[28] or other, even more unconventional mechanisms[30,31,32] Indeed the debate concerning the possible explanations for the PCAR spectra in the electron doped cuprates continues[25,29,30,31,32]. It has been suggested that both disorder[30] or the existence of antiferromagnetic order in the superconducting state of HTS materials[31,32] can explain the absence of nodes in the order parameter and therefore the absence of a zbcp in the conductance spectrum, both of which are highly possible in the system considered here. However, until directional STM or PCAR is performed on single crystal oxypnictides such questions must remain open. We note that it is encouraging in this regard that Nd and Sm based iron-arsenide single crystals have very recently been fabricated[33,34]. Nonetheless, the consistency of the PCAR studies of $2\Delta_0/k_BT_c \sim 3.5$ for two samples of very different composition and $T_c$ indicates that these samples share the same type of pairing mechanism and our observations of PCAR as a function of pressure suggests that the interpretation of a simple d-wave order parameter may be premature. Indeed the weak features at higher bias shown in figure 1 may be an indication of multiple gaps in this material[35,36]

Since drafting this manuscript we have become aware of a point contact study on $SmFeAsO_{0.85}F_{0.15}$ in which only one apparently nodeless gap is observed [37]. The data is in many ways consistent with our own observations. However only the availability of pure single crystals or epitaxial films will fully resolve the fundamental question associated with the symmetry of the order parameter.

**Figure Captions**

Figure 1: (a) PCAR as a function of temperature showing a pronounced zbcp. (b) The magnitude of the zbcp normalised to the conductance spectrum above $T_c$ as a function of temperature for two contacts A (■) and B (O). (c) Spectra at 39K for the different contacts shown in (b) showing different values of the zbcp, contact A (■) and contact B (O).

Figure 2: Point contact spectra with an Au tip at 4.2K at two different positions (a) and (b) as a function of contact pressure.

Figure 3: A Point contact spectrum at low tip pressure (T = 4.2K), with a fit (red line) to the BTK model showing fit parameters of $\Delta$ = 5.94 meV, Z = 0.44, smearing parameter, $\omega$ = 2.03 meV.

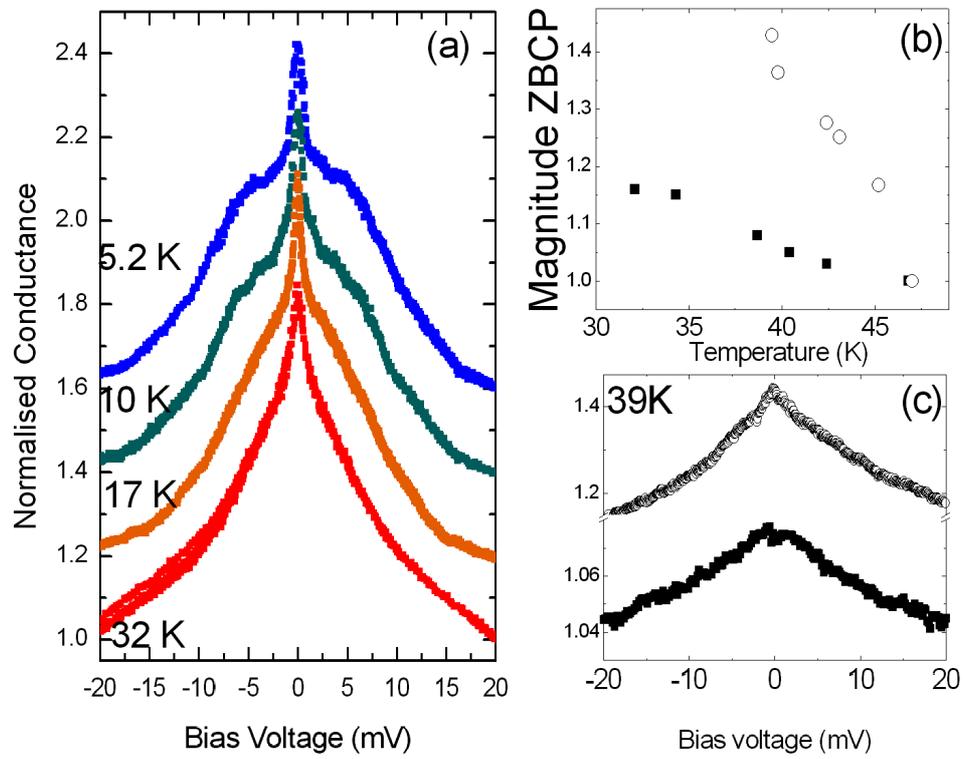

Figure 1, Yates et al

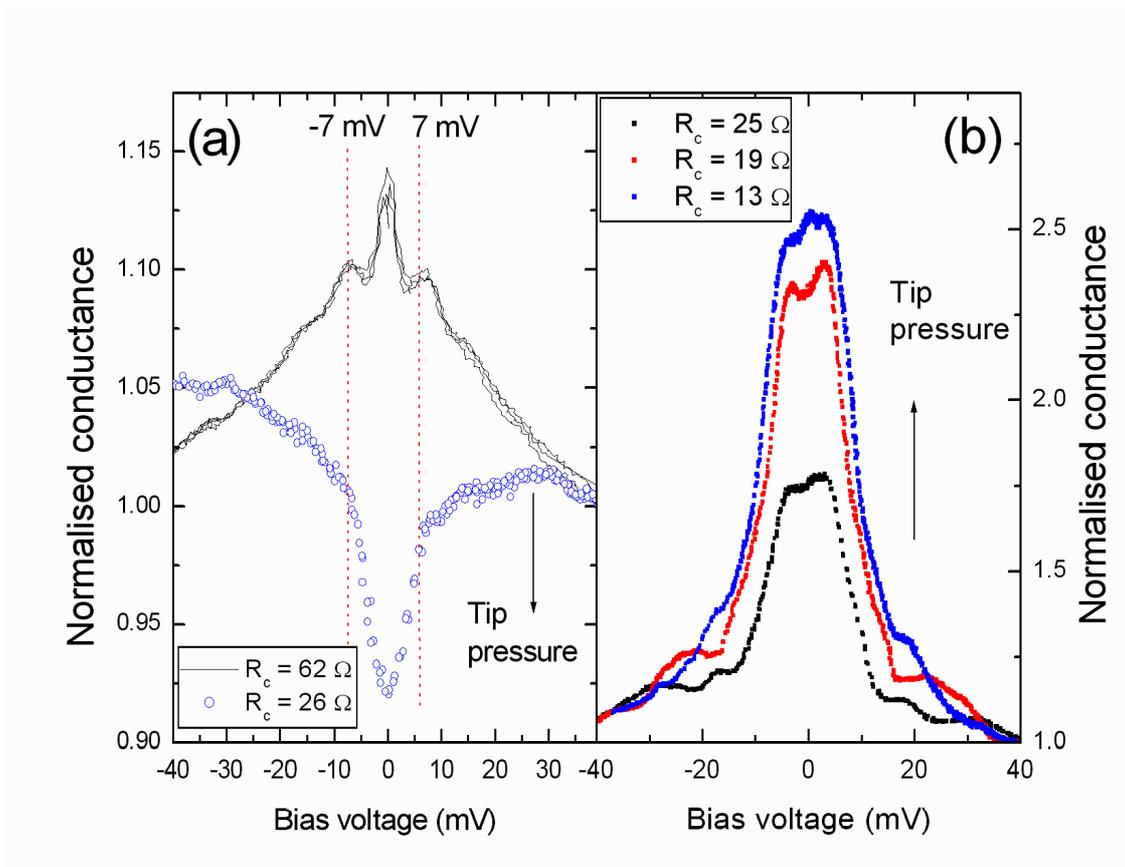

Figure 2, Yates et al

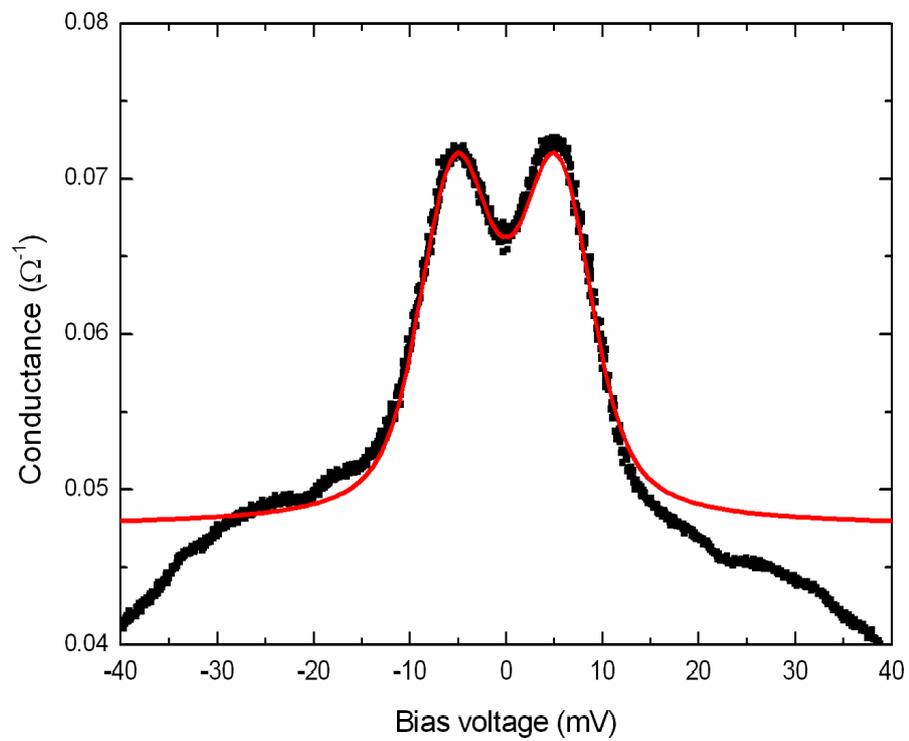

Figure 3, Yates et al